\documentclass[twocolumn]{aastex631}

\usepackage{pxfonts}

\usepackage{color}
\usepackage{hyperref}
\usepackage[shortlabels]{enumitem}
\usepackage{comment}    

\shorttitle{Unified rapid MT}
\shortauthors{Ivanova et al.}

\graphicspath{{./}{figures/}}

\accepted{June 5, 2024}

\begin{document}

\title{Unified Rapid Mass Transfer}

\author[0000-0001-6251-5315]{Natalia Ivanova}
\affiliation{Department of Physics, University of Alberta, Edmonton, T6G 2E7, Alberta, Canada}

\author{Surjakanta Kundu}
\affiliation{Department of Physics, University of Alberta, Edmonton, T6G 2E7, Alberta, Canada}
\affiliation{ Department of Physical Sciences, Indian Institute of Science Education and Research Kolkata, Mohanpur 741246, India}

\author[0000-0002-5414-243X]{Ali Pourmand}
\affiliation{Department of Physics, University of Alberta, Edmonton, T6G 2E7, Alberta, Canada}

\correspondingauthor{Natalia Ivanova}
\email{nata.ivanova@ualberta.ca}

\begin{abstract}
We present a method to obtain rapid mass loss rates in binary systems, specifically at the onset of MT episodes.
The method unifies atmospheric (underflow) and $L_1$ stream (overflow) mass rates in a single continuous procedure.
The method uses averaged 3D properties of the binaries, such as effective binary potential and effective binary acceleration, to both evolve the donor and obtain properties of the matter at the $L_1$ plane. In the case of underflow, we obtain atmospheric stratification.
Our method can be used for binaries with an extensive range of mass ratios, $0.01 \le q \le 100$, and can also be applied to hot donors. 
The considered examples show that the MT rates obtained with this revised formalism always differ from the optically thin and optically thick MT rates widely used during the computations of binary evolution.
\end{abstract}

\keywords{Multiple star evolution --- Binary stars --- Roche lobe overflow  }

\section{Introduction}

\label{sec:introduction}

Mass transfer (MT) is a type of interaction in the lives of close binaries when one star's outer layers become too close to the effective binary potential that passes through the first Lagrangian point, $L_1$, between the two binary companions. The surface of this equipotential is known as a Roche lobe. Proximity of the star to its Roche lobe allows the donor's outer layers to flow into the Roche lobe of its companion.

Stable MT can be a long-lived stage during which binaries appear for example as X-ray binaries, including low-mass X-ray binaries, high-mass X-ray binaries, cataclysmic variables, and more \cite[see review on mass-transferring systems in][]{2023pbse.book.....T}. A fast phase of  stable MT can be responsible for the formation of such exciting objects as Algols, stripped supernovae, and more. 

However, if the MT becomes dynamically unstable, the progenitor binary system may enter a Common Envelope (CE) phase. 
During a CE event,  one of the binary companions (accretor) orbits inside the envelope of another binary companion (donor).  This transitional evolutionary phase was proposed to explain cataclysmic variables \citep{1975PhDT.......165W} and V471 Tau \citep{1976IAUS...73...75P} \citep[for a more detailed history of the idea's origin and its development throughout last 50 years, see the review in][]{2020cee..book.....I}.
 Currently, a CE phase is considered to be a major transformational event responsible for forming the progenitors of the already mentioned X-ray binaries, and also progenitors of merging double neutron stars or double black holes.

A dynamical phase of a CE event starts when the donor overfills the equipotential that passes through the outer Lagrangian point. Understanding the initiation of a CE phase is crucial to answer CE outcomes. Specifically, how much mass was lost during the pre-CE mass loss, i.e., before the donor overfilled its outer Roche lobe?

There are several types of techniques to evolve the donor through the mass loss episode. One, often called ''implicit'', is to assume that the donor has to remain inside its Roche lobe. The MT rate is then found by taking away as much mass as needed to keep the donor exactly inside its Roche lobe. This method is excellent for finding smooth and continuous MT rates during stable MT. However, this approach is intrinsically inapplicable for modeling the start of a dynamically unstable MT leading to a CE event.

An opposite approach is when the donor is considered to not coincide precisely with its Roche lobe, and the flow is formed from the layers that exceed the Roche lobe.
The converging-diverging behavior of the effective binary potential in the $L_1$ neighborhood suggests that one may consider the $L_1$ neighborhood as an analogy of the de Laval nozzle, where the $L_1$ plane acts as a transition cross section between subsonic and transonic flows \citep{1967PASJ...19..564N}. Adding the assumption that streamlines of mass flow do not cross each other and follow the Bernoulli theorem,  \cite{1967PASJ...19..564N} introduced the way to find the rate of mass loss in the case of { isentropic} streams. While the applicability of the de Laval formalism for converging-diverging equipotentials was questioned \citep[for example, see][]{1975ApJ...198..383L}, it is still a standard and widely accepted approximation to find the mass loss rate.

The next comprehensive analysis of gas dynamics during the MT, though with the goal to understand the post $L_1$ ballistic stream trajectory, was developed in \cite{1975ApJ...198..383L}. The additional approximation discussed there and later used elsewhere was to consider the MT using an {isothermal} stream, and the use of a Taylor expansion for the converging-diverging behavior of the binary potential near $L_1$. 

Later, \citet{eggleton1983approximations} made three dimensional (3D) integrations of the Roche lobe and derived the now famous empirical formula for a volume equivalent Roche lobe radius.  \cite{mochnacki1984accurate}, also using 3D integrations but outside of the Roche lobe, provided additional quantities specifically related to binary potentials.

This set of approximations for a Roche lobe geometry led to the development of what is considered now to be the two most commonly used prescriptions for MT rates, both of which are functions of how much the donor overfills or underfills its Roche lobe.

The Roche lobe ``underfilling'' MT occurs since a star has no rigid surface at its photosphere, and the donor's material extends beyond what is usually called the star's radius. Then, one can find how much mass  can be lost from the donor's atmosphere before the donor's radius formally becomes equal to its Roche lobe volume-equivalent radius. The detailed consideration of this prescription using the isothermal stream consideration for cold (low-effective temperature) donors was done by \cite{Ritter1988}. Importantly, it used an empirical equation for the potential that is based on \cite{mochnacki1984accurate} 3D integrations and is valid for mass ratios (the ratio of the donor mass to the mass of its companion) $0.1\la q \la 3$. The surface temperature of the donor should be cold enough so that there is no significant radiation pressure. We will refer to this formalism as Original Ritter ({\bf OR}) formalism. The mass loss considered this way is the optically thin MT or atmospheric overflow. 

Roche lobe overflow (RLOF) can also be considered in a limit of an  optically thick stream formalism \citep{1990A&A...236..385K}. We will refer to this formalism as {\bf KR} formalism. It uses a combination of an adiabatic stream, a second-order Taylor expansion near $L_1$ for potential, and again a matter-dominated approximation for the total pressure.

One has to choose which approach to use in their calculation, optically thin or optically thick, based on other physics of the considered binary system. {\bf OR} can be used for both underflow and overflow, but it is problematic to justify that a thick $L_1$ stream is optically thin. 
On the other hand, {\bf KR} by design can only find  MT rates in case of an overflow, and hence using {\bf KR} usually implies that {\bf OR} is used while the system is still in underflow.
We note, however, that the physical assumptions that are used to derive optically thin and optically thick mass loss rates differ, and by design, there is no continuity between the two regimes. 

The donors that enter the CE event do not have to be as cold as those in cataclysmic variables, for which {\bf OR} and {\bf KR} were derived. Further, a dynamical instability of the MT is often the consequence when the mass ratio of the donor to its companion is large, $q>3$ \footnote{There is no unique value of the critical mass ratio $q_{\rm crit}$ as it is highly dependent on the evolutionary stage of the donor, the type of its companion, and the assumed mode of the MT. As guidance,  $q>q_{\rm crit} \approx 3.5$ would most likely lead to dynamically unstable MT for radiative donors, $q_{\rm crit} \approx 2$ for low and intermediate-mass convective donors, and $q_{\rm crit} \approx 8$ for very massive donors. For more details, see the review in \cite{2020cee..book.....I}.}.  
The dynamically unstable MT also proceeds with a significant degree of overflow. That is beyond the applicability of the empirical approximation integrated in  {\bf OR} formalism and the second-order Taylor expansion used in {\bf KR}. An additional degree of inconsistency lies in using the parametrization of the binary effective potential for the stream condition with a star that is not evolved in the same potential. 

Recently, \cite{2023ApJ...952..126P} (Paper I) have described a method to obtain 3D averaged stellar properties for a star inside a binary potential, from the center of the donor to the equipotential passing through the donor's outer Lagrangian point, as well as the properties at the $L_1$ plane, for an extensive range of mass ratios, $10^{-5}\le q \le 10^5$. The obtained tables now enable integrating a star's evolution in a binary potential. The next step is to obtain the mass loss rates which use the average 3D properties.

In this paper,  we first re-derive the atmospheric overflow and RLOF formalisms, see \S~2. The derived unified formalism represents the continuous approach. In \S~3, we compare the revised unified formalism with {\bf OR} and {\rm KR} formalisms. In \S~4, we evaluate the trends of the impact of the new formalism on the mass transferring system.

\section{Method description}

\subsection{Roche lobe overflow}

For MT calculations when the donor overfills its Roche lobe, we use, as the foundation, the optically thick stream formalism developed by our group, \cite{2015MNRAS.449.4415P}, updated here to take into account the averaging of the binary properties from 3D to 1D as published in  Paper I.

As in \cite{2015MNRAS.449.4415P}, we obtain the MT rate by integrating the mass flow over a ``nozzle'' cross-section that is taken on the plane perpendicular to the line connecting the centers of the two stars and passing through the L1 point: 

\begin{equation}
\dot M_{\rm RLOF} = \oiint_{L_1 nozzle} \rho_{L1}(x,y) c_{s,L1}(x,y) dA \ .
\label{eq:mrlofxy}
\end{equation}

\noindent Here $(x,y)$ are the coordinates on the $L_1$ plane,  $\rho_{L1}(x,y)$  is the density of the donor's material as it reaches the $L_1$ plane, $c_{s,L1}(x,y)$ is the sonic velocity of the gas (we adopt the conventional nozzle approximation, in which the speed with which gas passes the nozzle is equal to its local sonic velocity), and $dA$ is the area element of the nozzle cross-section at the $L_1$-plane. 

Each location $(x,y)$  on the $L_1$ plane has a value of the effective binary potential $\Phi_l(x,y)$. A group of points $(x,y)$ on the $L1$ plane that has the same value of $\Phi_l(x,y)$ forms a closed loop. The closed loop (circuit)  containing all the points that have a value of the effective potential $\Phi_l$ is ${\mathcal{L}}({\Phi_l})$. 
All thermodynamic quantities, for example, density or sonic velocity, are the same along each circuit ${\mathcal{L}}({\Phi_l})$.
For each $\Phi_l$, we approximate the circuit $\mathcal{L}({\Phi_l})$ with an ellipse $\mathcal{E}({\Phi_l})$ on $L_1$ plane. Each ellipse $\mathcal{E}({\Phi_l})$ can be described by its semi-major axis, $r_l$, and its semi-minor axis. The semi-major and semi-minor axes of the ellipse, as a function of $\Phi$, are tabulated values from Paper I, and $\mathcal{E}({\Phi_l})$ is the ellipse circumference.
The nozzle for this stream is limited by the effective potential of the donor's photosphere  $\Phi_{\rm ph}$, which corresponds to the ellipse  $\mathcal{E}_{\rm ph}\equiv\mathcal{E}({\Phi_{\rm ph}})$ with the associated semi-major axis $r_{\rm ph}$. We will simplify notations, stating  that for a quantity $\cal{Q}$ at the equipotential loop determined by $\Phi(r_l)$, the equivalency in notations is implied  as follows:

$${\cal{Q}}(\Phi(r_l))\equiv {\cal{Q}}(r_l)\equiv {\cal{Q}}(\Phi_l) \ . $$

\noindent 
Equation \ref{eq:mrlofxy} can be rewritten as

\begin{equation}
\dot M_{\rm RLOF} = \int_{0}^{r_{\rm ph}} \rho_{L1}(r_l) \ c_{s,L1}(r_l) \ \mathcal{E}(r_l) dr_l \ .
\end{equation}

We can find at which volume equivalent radius $r_i$ the donor would experience the same effective binary potential: for each circuit on $L_1$ plane described by $r_l$, we find the volume-equivalent radius of the donor $r_i$ such that $\Phi(r_i) =\Phi(r_l)$. To find $r_i$, we use tabulated values from Paper I. The main improvement, as compared to \cite{2015MNRAS.449.4415P}, is that now we can precisely relate the properties of a gas at the $L_1$ plane with the same properties in the equipotential shells inside the donor.

Following \cite{2015MNRAS.449.4415P},  and assuming that each streamline is adiabatic (while the flow through the $L_1$-plane itself is not isentropic), we obtain $\rho_{L1}(\Phi(r_l))\equiv \rho_{L1}(r_l)$ and $c_{s, L1}(r_l)$ from solving the Bernoulli equation for streamlines along the equipotentials:

\begin{equation}
\frac{P(r_i)}{\rho(r_i)} + u(r_i) = \frac{c_{s,L1}^2(r_l)}{2} + \frac{P_{L1}(r_l)}{\rho_{L1}(r_l)} + u_{L1}(r_l) \ . 
\end{equation}

\noindent \cite{2015MNRAS.449.4415P} have demonstrated that using the polytropic equation of state, instead of the full tabulated equation of state as provided by {\tt MESA}, results in a difference for the final stream mass loss within 4 percent. In this work, we chose to use the polytropic equation of state to speed up the calculations. In that case, the density and sonic velocity on the $L_1$ plane can be found as

\begin{eqnarray}
        \rho_{L1}(r_l) &=& \left (\frac{2}{\gamma_i+1} \right )^\frac{1}{\gamma_i-1}\rho(r_i) \nonumber \\
        c_{s,L1}^2(r_l) &=& \gamma_i \left (\rho_{L1}(r_l)\right )^{\gamma_i-1} \frac{P(r_i)}{\rho_{L1}(r_i)^{\gamma_i}}
\end{eqnarray}

Here $\gamma_i$ is $\gamma\equiv\Gamma_1 = \left ( \partial \ln P / \partial \ln \rho \right )_{\rm ad}$ in the donor at $r_i$.

\subsection{Atmospheric overflow}

\label{sec:atmooverflow}

For mass loss from the atmosphere, we use  Ritter's formalism \citep{Ritter1988}, but strongly modified.
To be specific, the original Ritter's formalism assumes that both the atmosphere and the stream are isothermal, and then applies a number of approximations to find the mass loss rate.
In what follows, we will discuss the isothermal and adiabatic assumptions and detail which approximations can be replaced with exact integrations. 

We will start with the overall setup. The original Ritter's formalism adopts that both the atmosphere of the donor (to be specific, all the matter above the photosphere) and the stream, while it travels to the $L_1$ plane, are kept isothermal, and the material reaches the sonic speed at the nozzle ($L_1$ plane). Since the stream is isothermal due to external radiation, only the  form of the Bernoulli equation derived from the momentum conservation can be used:

$$
\frac{1}{2} v^2 + \int \frac{dP}{\rho } +\Phi = {\rm const}
$$

\noindent To find the solution, {\bf OR} evaluates the Bernoulli constant at the photosphere and at the $L_1$ plane.
This implies that the streamlines are meant to start at the photosphere and then are expected to arrive at the $L_1$ plane. The external energy source is mandatory here, as without an external energy source, a flow of matter from a deep potential well at the photosphere to the $L_1$ plane while accelerating to sonic velocity and keeping the same temperature is against energy conservation. Using this assumption, the density at $L_1$ in {\bf OR} is  

\begin{equation}
\rho_{\rm L1}^{\rm is}(\Phi_{L1}) = \frac{1}{\sqrt e}\rho_{\rm ph} \exp \left (- \frac{\Phi_{L1}-\Phi_{\rm ph}}{c_{s,\rm ph}^2} \right )
\label{eq:is_rho_l1}
\end{equation}

\noindent  Here $\Phi_{L1}$ is the effective binary potential at $L_1$ and $\Phi_{\rm ph}$ is the effective binary potential at the photosphere, $\rho_{\rm ph}$ is the density of the photosphere, and $c_{s,\rm ph}$ is the sonic velocity in the photosphere, atmosphere and $L_1$ plane. The equation above adopts the use of a standard set of assumptions describing an isothermal astrophysical flow: $P\propto \rho$, and isothermal sonic velocity $c_{\rm s}^{\rm is} = \sqrt{P/\rho} = {\rm const}$.

We note that while the original setup makes the stream originate from the photosphere and end at the $L_1$ plane, the setup is effectively equivalent to  two separate assumptions: a) the atmosphere is in hydrostatic equilibrium and is isothermal due to long-term irradiation by the donor; b) matter leaves the atmosphere along the equipotentials and remains isothermal as it travels to $L_1$.
 
Having the two separate assumptions adds more insight. First, the assumption for the donor's atmosphere to be in  isothermal equilibrium is a frequent assumption for a stellar atmosphere (although see \S\ref{sec:atmo}). Second, matter streaming along the equipotentials is more self-consistent and aligns with how RLOF mass loss is seen\footnote{The ratio of Coriolis acceleration to the effective binary acceleration on $L_1$ plane is comparable to the ratio of sonic velocity at $L_1$ plane to the product of binary angular velocity and the distance to $L_1$ plane from the center of the donor. The described common assumption on the stream being aligned with equipotentials implies that Coriolis acceleration is chosen to be ignored, for being a second-order correction, see \cite{1975ApJ...198..383L}.}. However, unlike the assumption for the hydrostatic atmosphere to be isothermal due to long-term donor irradiation, the stream, as it starts to flow, does not have to remain isothermal, especially if the considered MT episode is dynamically unstable. In what follows, we will consider separately if the stream material remains isothermal as it flows along an equipotential, or adiabatic. 

Let us start from the assumption that the atmosphere is isothermal and is in hydrostatic equilibrium. Then, pressure and density in the atmosphere at the equipotential $\Phi_i$ that corresponds to the volume equivalent radius $r_i$ are

\begin{eqnarray}
P_{\rm at}(\Phi_i) &=&  P_{\rm ph} \exp \left (- \frac{\Phi_{i}-\Phi_{\rm ph}}{c_{s,\rm ph}^2} \right ) \nonumber \\
 \rho_{\rm at}(\Phi_i) &=& \rho_{\rm ph} \exp \left (-\frac{\Phi_{i}-\Phi_{\rm ph}}{c_{s,\rm ph}^2} \right )
\end{eqnarray}

A further assumption, for the stream to remain isothermal while it flows along the equipotential, leads to the same expression for the density at $L_1$ as in Equation~\ref{eq:is_rho_l1}. However, as we speak now about {\it different} equipotentials in the atmosphere, we can also  find the density at any location in the $L_1$ plane:

\begin{eqnarray}
\rho_{L1}^{\rm is}(\Phi_i) &=&  \frac{1}{\sqrt{e}} \rho_{\rm ph} \exp \left (- \frac{\Phi_{i}-\Phi_{\rm ph}}{c_{s,\rm ph}^2} \right ) \nonumber \\
c^{\rm is}_{s,L1}(\Phi_i) &=&c_{s, \rm  ph} = \sqrt{\frac{P_{\rm ph}}{\rho_{\rm ph}}}
\end{eqnarray}

\noindent In this scenario, the sonic speed remains constant 
along the stream 
and has, at the $L_1$ plane, the same value as at the photosphere. The combination of isothermal atmosphere and isothermal stream recovers Ritter's values at $L_1$.

If, instead, the matter is assumed to evolve adiabatically as it travels along the equipotential, while it is still departing from the isothermal atmosphere, 

\begin{eqnarray}
\rho_{L1}^{\rm ad}(\Phi_i) &=&  
\left ( \frac{2}{\gamma_i+1} \right )^{\frac{1}{\gamma_i-1}}\rho_{\rm ph} \exp \left (- \frac{\Phi_{i}-\Phi_{\rm ph}}{c_{s,\rm ph}^2} \right ) \nonumber \\
 c_{s,L1}^{\rm ad }(\Phi_i) &=&  \sqrt{\frac{2}{\gamma_i +1 }}c_{s,\rm ph}
\end{eqnarray}

\noindent Here $\gamma_i$ is $\gamma\equiv \Gamma_1$ at the corresponding equipotential level of the atmosphere, and the sonic velocity of the atmosphere here is the same quantity as was used to make the isothermal atmosphere,  $c_{s,\rm ph}=\sqrt{P_{\rm ph}/\rho_{\rm ph}}$. 

 Note that  integration over the whole $L_1$ plane is implied, with $r_{\rm out}$ reaching the outer Lagrangian equipotential in the $L_1$ plane. The mass loss rate can be obtained for the isothermal or adiabatic scenarios using the corresponding density and sonic velocity expressions. As at each location in the $L_1$ plane, the product of $\rho_{L1}^{\rm ad}(\Phi_i)c^{\rm is}_{s,L1}(\Phi_i)$ and the product of $\rho_{L1}^{\rm ad}(\Phi_i) c_{s,L1}^{\rm ad }(\Phi_i)$ differ only by a constant factor of about one, the integrated MT rates from an isothermal atmosphere with either an isothermal or adiabatic stream will be about the same.  For example, for $\gamma=5/3$, an isothermal stream from an isothermal atmosphere will result in MT rates that are about 1.078 times those assuming an adiabatic stream from an isothermal atmosphere, and for $\gamma=4/3$ the ratio is 1.04.

In addition to isothermal and adiabatic atmospheres, one can construct a simple atmosphere model and then apply the isothermal or adiabatic Bernoulli equation for each atmospheric layer to obtain values of the density $\rho_{L1}(r_l)$ and sonic velocity $c_{s,L1}(r_l)$ as functions of the position (or the effective potential) at $L_1$ plane. There is no analytical solution in this case. Further, the construction of the atmosphere can proceed with several approximations, which we will list here:

\begin{enumerate}
\item {\bf EG.} This is a classic Eddington grey (EG) atmosphere in which both the opacity of the photosphere $\kappa_{\rm ph}$ and the gravitational acceleration $g_{\rm ph} =(d\Phi/dr)_{\rm ph})$ are kept constant through the atmosphere \citep[see, e.g., \S4.3 in][]{2004sipp.book.....H}.
\item {\bf EG$s$.} This atmospheric model considers that the atmosphere can be extended, and hence, its geometry is not fully plane-parallel; the radiative energy transfer equation must be solved in spherical geometry in this case. \cite{1969AcA....19....1P}  noted that extended atmospheres have an intrinsically large sensitivity to the adopted (initially arbitrary) value of the temperature at the true surface (where density becomes zero). To resolve this numerical nuisance, a further approximation must be made. For example, \cite{1969AcA....19....1P} proposed adding an approximate slope in the equation for the temperature gradient (see their Equation 7). 
\item {\bf EG$\Phi$.} In this atmospheric model, the effective gravitational acceleration is not constant.
\item {\bf EG$\Phi\kappa$.} In this variation, both the opacities and the effective gravitational acceleration are functions of the radius.
\end{enumerate}

Please note that only the {\bf EG} atmosphere would be called an Eddington grey atmosphere by default in the literature.

The total mass loss rate, in case of only atmospheric outflow, is 

\begin{equation}
\dot M_{\rm at} = \int_0^{r_{\rm out}} \rho_{at}(r_l) \ c_{s,\rm at} (r_l) \ \mathcal{E}(r_l) dr_l
\end{equation}

\subsection{The total mass loss rate.}

Now, we can obtain the total mass loss rate when the donor overfills the Roche lobe while keeping an atmospheric outflow. The combined mass loss rate is important for transitioning from the Roche lobe underfilling (atmospheric) outflow to the Roche lobe overfilling outflow.

\begin{eqnarray}
\dot M_{\rm tot} &=& \int_{0}^{r_{\rm ph}} 
\rho_{L1}(r_l) \ c_{s,L1}(r_l) \ \mathcal{E}(r_l) dr_l  \nonumber \\
&+& \int_{r_{\rm ph}}^{r_{\rm out}} \rho_{at}(r_l) \ 
c_{s,\rm at} (r_l) \mathcal{E}(r_l) dr_l
\end{eqnarray}

\subsection{Fast mass removal.}

Currently, when only the total mass loss rate is provided by approximate prescriptions, the mass loss in stellar codes can be applied only to a donor as a whole. Essentially, during the numerical iterations for a stellar model at the next timestep, the atmosphere is removed and reconstructed right above the mass shell that previously was deep inside. That motivated the idea that the fast mass loss has to be adiabatic, as the deep layers would have no time to lose energy during fast mass loss. Accordingly, studies of the stability of the MT are based on the adiabatic response of the envelope. The deviation from adiabaticity so far was linked only to the reconstruction of the superadiabatic layer in the very outer layers of the envelope \citep{2015MNRAS.449.4415P}.

With our mass loss formalism, we can obtain the mass loss rate for each individual mass shell inside the donor.  
This only plays a role during RLOF mass loss, where the matter is streaming directly from the equipotential inside the donor below the photosphere.

Consider a mass shell $m_i$ with a mass $dm_i$, thickness $dr_i$, located at the volume-equivalent radius inside the donor $r_i$.
With $\Phi(r_l)=\Phi(r_i)$, $r_i$ corresponds to the semi-major axis $r_l$ at the $L_1$ plane, while the thickness $dr_i$ projects to a thickness $dr_l$ at $L_1$ plane. The mass loss from the $i$ mass shell is 

\begin{equation}
\dot m(r_i) = 
\int_{r_l}^{r_l +dr_l}
\rho_{L1}(r_l) \ c_{s,L1}(r_l) \ \mathcal{E}(r_l) dr_l
\end{equation}

\hypertarget{link_example}{ For mass shells with a small thickness $dr_i$ } and correspondingly small $dr_l$,

\begin{equation}
\dot m(r_i) = - 
\rho_{L1}(r_l) \ c_{s,L1}(r_l) \ \mathcal{E}(r_l) dr_l \ .
\end{equation}

However, from the point of computation time, removing mass directly from the mass shells can be useful only when the mass loss rate approaches dynamical—only then can a fraction of a mass be removed from each mass shell during a timestep. However, another situation can occur before the mass loss rate becomes dynamic.

Let us introduce ``sonic'' mass loss from the photosphere of a 1D star as the following:

\begin{equation}
\dot M_{\rm sonic} = 4 \pi r_{\rm ph}^2 c_{s,\rm ph} \rho_{\rm ph} \ .
\end{equation}

During a significant Roche lobe overflow, the sonic speed and density near $L_1$ can greatly exceed those near the photosphere. As a result, the mass loss rate found by integrating over the $L_1$ neighborhood can exceed $\dot M_{\rm sonic}$. Numerically, if the mass loss is applied only to the outer layer, the code must recover the stellar model from the supersonic removal of it. Hence, it leads numerically to adiabatic mass removal of the surface superadiabic layer. However, in a real star, the outer layers do not leave the star at supersonic speed. Instead, the mass loss takes place through the range of mass shells, and it is everywhere moving at subsonic speed on average.
We know which mass shells are attached to the $L_1$ plane at the moment $t$, and we 
can predict which mass shells will be in contact with the L1 plane during the next timestep. We then remove the mass that has to be lost from those mass shells. We term this as predictive mass shell removal.

Both dynamical mass shell removal and predictive mass shell removal are computationally intensive. During the mass transfer calculation during the binary evolution, we first use {\tt MESA} standard photospheric removal. If the mass loss removal approaches sonic, we switch to predictive shell mass removal. If the mass loss approaches a dynamical timescale rate, we use dynamical mass loss removal. However, the comparison with {\bf KR} mass loss can only be made using {\tt MESA} standard photospheric removal.

\section{Comparison of the formalisms.}

To test our unified MT formalism, we use the Module for Experiments in Stellar Astrophysics package, {\tt MESA} (\cite{2011MESA}, \cite{2013MESA}, \cite{Paxton2015}, \cite{2018MESA}, \cite{2019MESA}, \cite{2023MESA}), revision 23.05.1. {\tt MESA} has integrated prescriptions for mass loss. The option "Ritter" uses the {\bf OR} formalism, and the option ''Kolb'' uses {\bf KR} formalism. 

For our formalism, in addition to different mass loss rates, we evolve the donors using the effective binary acceleration in a corotating frame, using the subroutine published in Paper I. To provide smoother MT calculations, we replaced the tables provided in Paper I with more refined ones: the new tables are obtained for the mass ratios $-2\le \lg_{10} q \le 2$ with steps of 0.01 in $\lg_{10} q$, and 500 points resolution between $L_1$ and $L_{\rm out}$ potential instead of 100 points in the original tables.
Effective accelerations and volume-equivalent radii for all equipotentials were found numerically as in Paper I, with the precision $10^{-5}$ of its value or better. 
In all the cases (testing {\bf OR}, {\bf KR}, or our formalism), the simulations stopped when the donor reached $R_{\rm lout}$, which is the volume-equivalent radius of the outer Lagrangian equipotential.

\begin{figure*}
    \centering

    \includegraphics[width=0.49\textwidth]{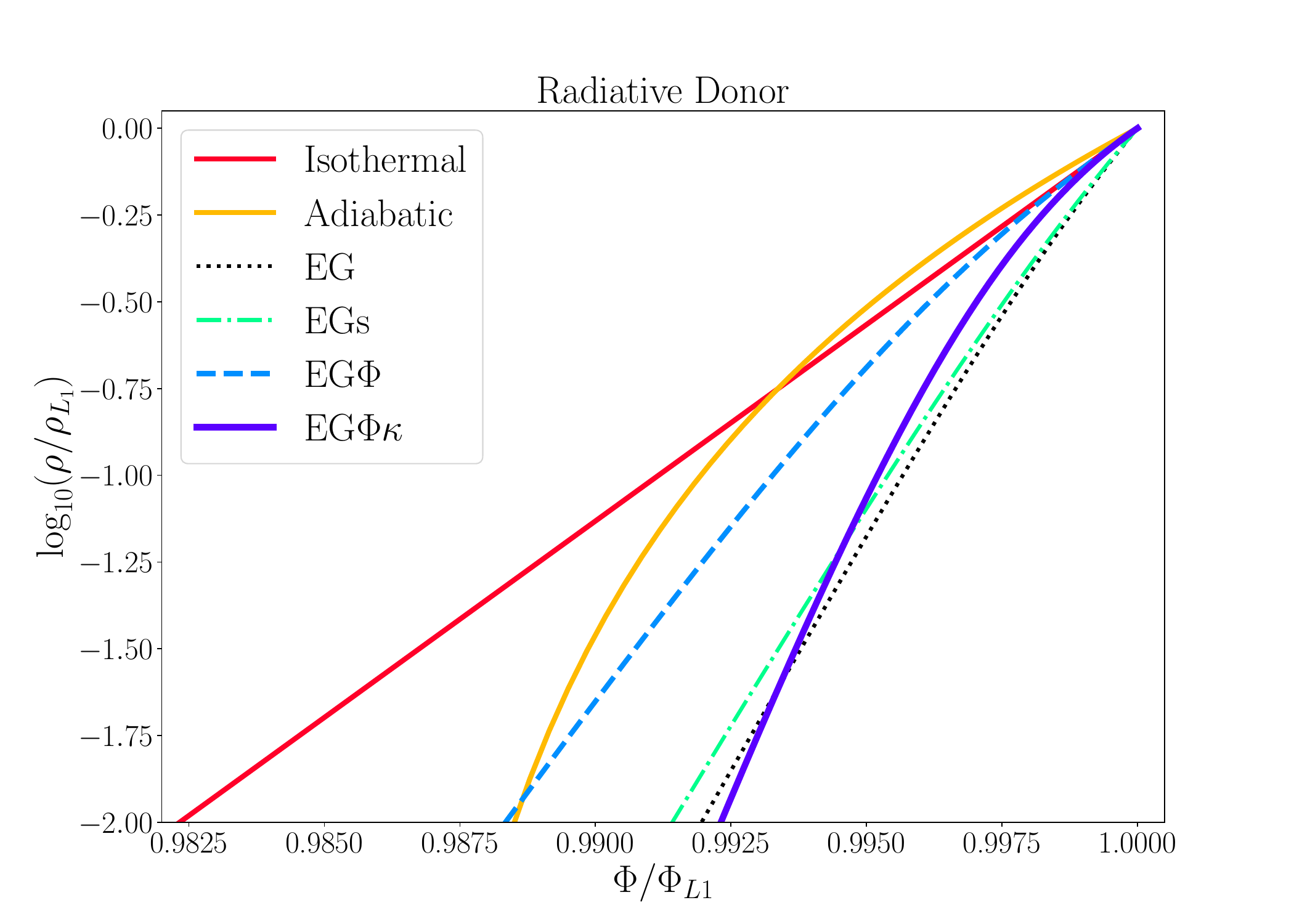}
    \includegraphics[width=0.49\textwidth]{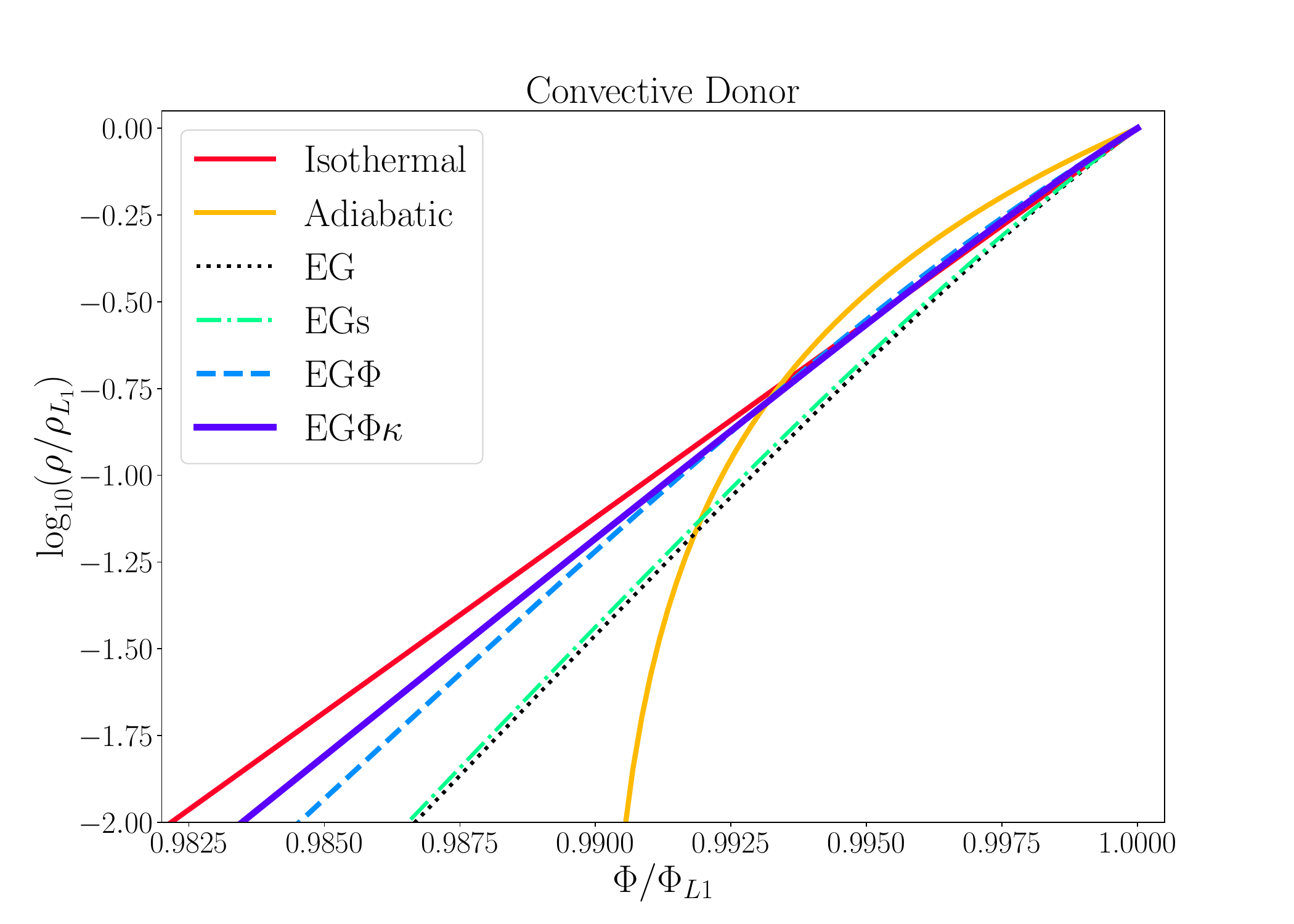}
    \caption{Relative density profiles for several atmospheric models in a radiative (the left panel) and a convective donor (the right panel), as a function of equipotential. The atmospheric models as described in \S\ref{sec:atmooverflow}, specifically: {\bf EG} is standard Eddington grey, {\bf EG$\Phi$} is the model where the effective binary acceleration is non-constant, {\bf EG$\Phi\kappa$} is the model where both the effective binary acceleration and the opacities are non-constant, {\bf EGs} is the default {\tt MESA} model with a correction for non-planar atmosphere, {\bf Isothermal} is an isothermal atmosphere as in Ritter's formalism, and ``Adiabatic'' is when the entropy of all atmospheric layers is the same as that of the photosphere. The initial mass of the donor is $30$~$M_\odot$, the mass ratio is 6. The atmospheres were obtained when the donor's radius is $22.3$~$R_\odot$ (its effective temperature is $27200$~$K$) and $1340$~$R_\odot$ (its effective temperature is $3580$~$K$).}
    \label{fig:atmos}
\end{figure*}

\subsection{Atmospheres}

\label{sec:atmo}

Different atmospheric models (as described in \S\ref{sec:atmooverflow}) result in different density and sonic velocity profiles at the $L_1$ plane. To demonstrate the difference, we constructed different atmospheres. In each of the simulations, the photosphere is located at the $L_1$ equipotential when we constructed the profiles.

To be exact, {\bf EGs} is the self-consistent atmospheric model that can be obtained with {\tt MESA}, which is  termed there ''Eddington grey''. 
For the other models, we build the density, pressure, and temperature profiles in the neighborhood of the $L_1$ equipotential using the photospheric values of density, temperature, and pressure provided by {\tt MESA} as the initial condition. Those models for fluid stratification do not extend to $\tau=0$ and do not provide feedback on photospheric conditions.

The resulting density profiles, as functions of the relative effective potential, are shown in Figure \ref{fig:atmos}.

The comparison of isothermal and adiabatic atmospheres shows that the isothermal atmosphere has a more shallow power-law density decrease at large distances, and hence extends further. The adiabatic atmosphere has a stronger core  -- i.e., its density near the $L_1$-neighborhood remains larger -- and then it abruptly drops.
We also note that an isothermal atmosphere, while being built using the assumption that the $P/\rho$ ratio remains constant, does not result in an atmosphere that is truly isothermal. For example, as density drops by two orders of magnitude in Figure \ref{fig:atmos}, the temperature obtained from the full equation of state drops from 3580K to 3190K for the convective donor (and would drop to 2380K if density drops by three orders of magnitude). The reason for a drop in temperature is that in the atmosphere, a significant fraction of pressure is provided by radiation, with all pressure being provided by radiation at the ``true'' surface ($\rho=0$). It serves as a reminder that employing the equation for an ideal gas in the classic limit for the atmosphere, as is adopted in {\bf OR} formalism, can be inherently inconsistent, and its utilization is best approached with caution.

The comparison of the EG-based models shows that, at least for the tested models, the difference between the standard (plane-parallel) atmosphere (model {\bf EG}) and the one that introduces a correction for geometry (model {\bf EGs}) is insignificant. Varying opacities for the above models do not provide a significant effect either (not shown in Figure \ref{fig:atmos}, as it blends with the EGs model). The most significant effect on the density profile is when the effective acceleration is a function of the effective binary potential (model EG$\Phi$). Finally, an additional correction to that model is provided by using opacities that are a function of the local temperature and density (model EG$\Phi\kappa$). EG$\Phi\kappa$ behavior has some adiabatic properties when very close to $L_1$, and resembles an isothermal power law at  larger distances. Interestingly, for a convective donor, {\bf EG$\Phi\kappa$} results in the most ``isothermal'' stratification among all considered models, in the sense that for the given drop in density by 100 times from the photospheric value, the decline in temperature is the smallest -- it drops to 3290K. For a radiative donor, the {\bf EG$\Phi\kappa$} atmosphere resembles most the {\bf EG} atmosphere.

The latter model, {\bf EG$\Phi\kappa$}, is one that we adopt in what follows as the default atmospheric model for obtaining atmospheric overflow properties, while the isothermal and adiabatic models are used for some comparisons.

\begin{figure*}
    \centering
    \includegraphics[width=0.5\textwidth]{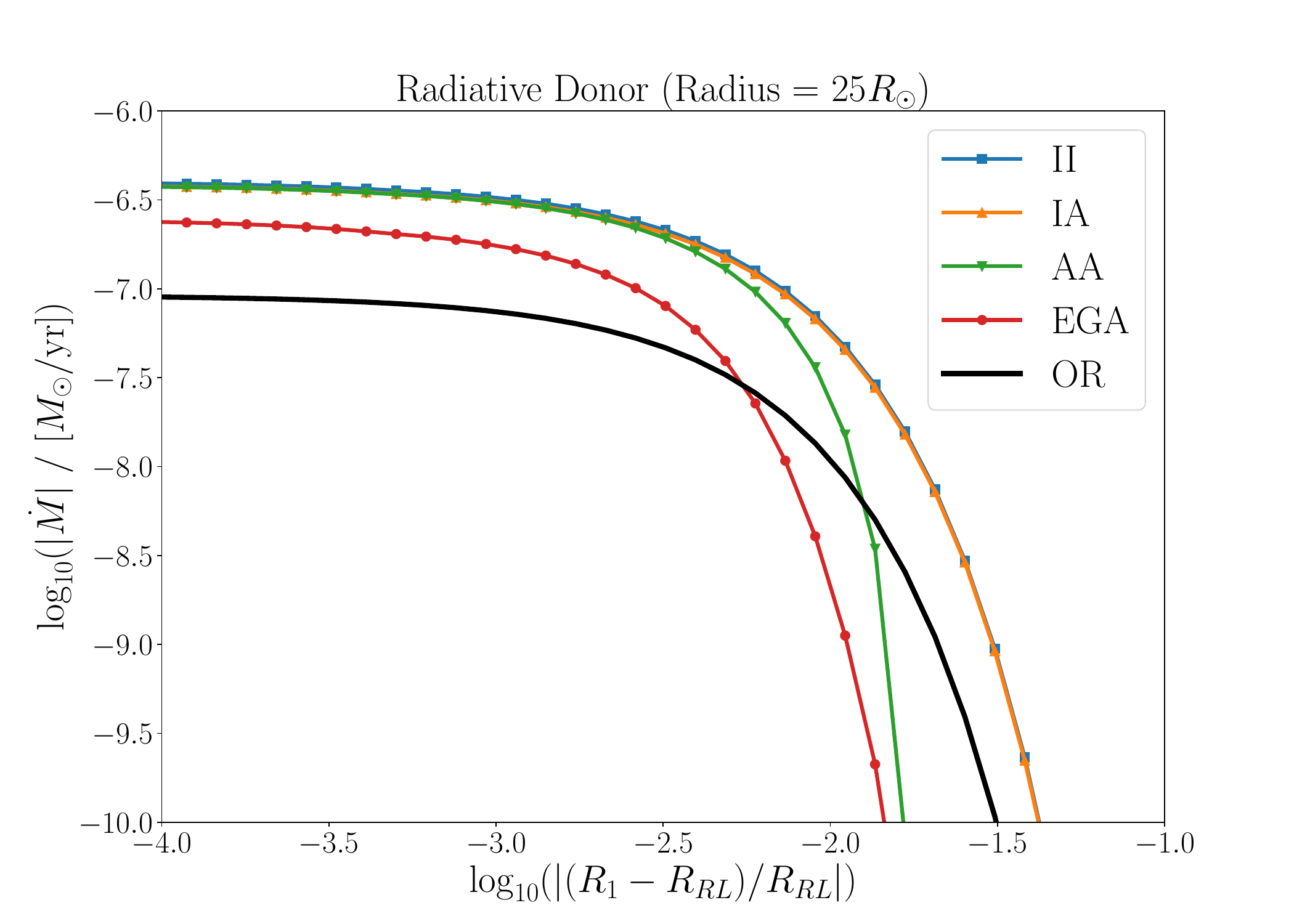}\includegraphics[width=0.5\textwidth]{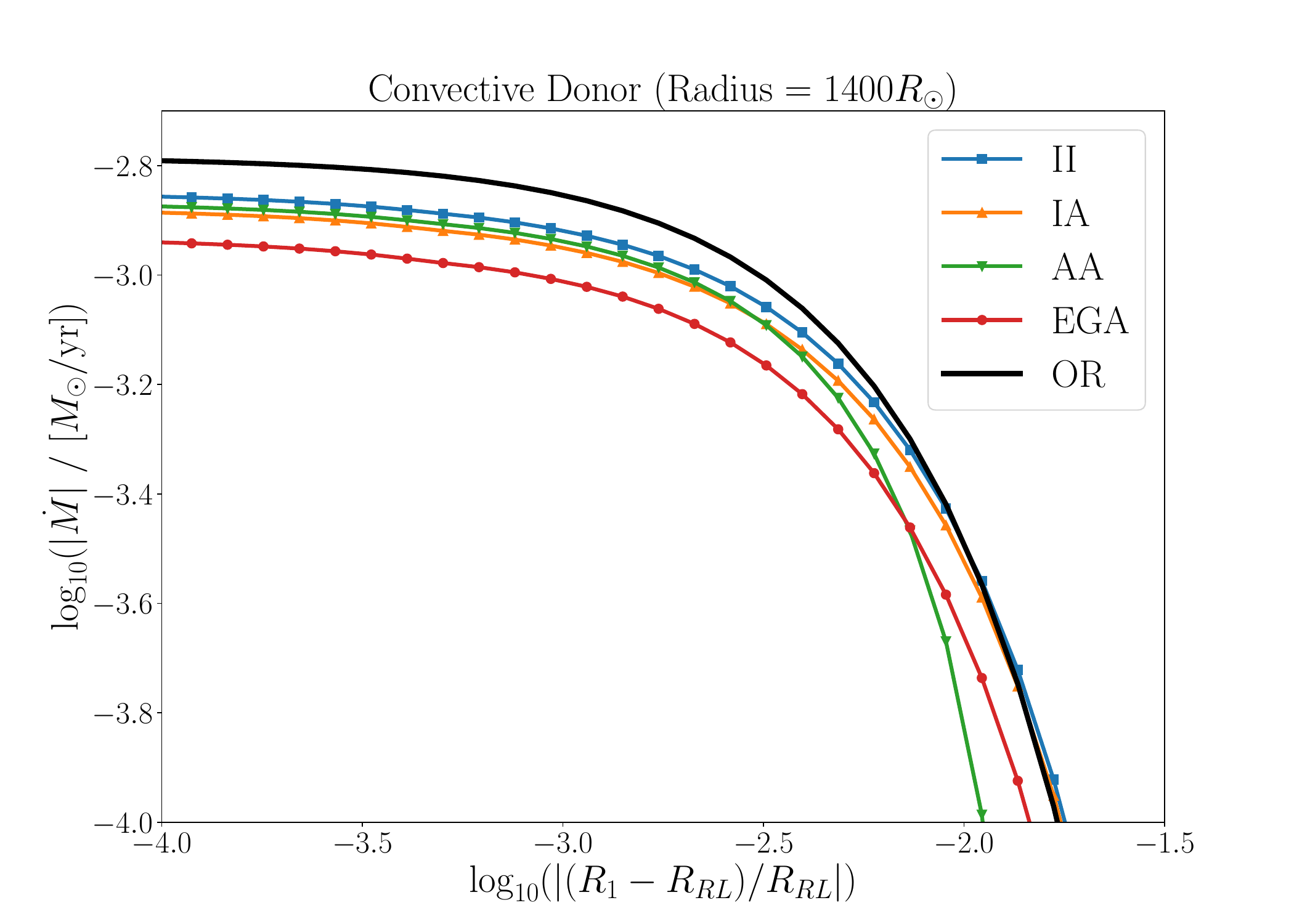}
    \caption{Atmospheric mass loss rates as a function of distance to the Roche lobe. A radiative donor is shown in the left panel, and a convective donor is shown in the right panel.  The models are described in \S\ref{sec:compatmooverflow}, specifically: {\bf II} means Isothermal atmosphere and Isothermal stream, {\bf IA} means Isothermal atmosphere and Adiabatic stream (almost identical to II in the left panel),  {\bf AA} means Adiabatic atmosphere and Adiabatic stream, {\bf EGA} means {\bf EG$\Phi\kappa$} atmosphere and Adiabatic stream, and {\bf OR} stands for Original Ritter formalism. The initial mass of the donor is $30$~$M_\odot$.}
    \label{fig:mt_at}
\end{figure*}

\subsection{Atmospheric overflow}

\label{sec:compatmooverflow}

Here, we will compare the mass loss rates for atmospheric overflow with the following assumptions:

\begin{itemize}

\item {\bf OR}: Original Ritter formalism 

\item {\bf II}: Isothermal atmosphere and Isothermal stream

\item {\bf IA}: Isothermal atmosphere and Adiabatic stream

\item {\bf AA}: Adiabatic atmosphere and Adiabatic stream

\item {\bf EGA}: EG$\Phi\kappa$ atmosphere and Adiabatic stream

\end{itemize}

In the Ritter formalism, the total mass rate is obtained as

\begin{equation}
    \dot{M}_{\rm at}^{\rm Ritter} = \rho_{L1} c_{s} Q \ .
    \label{eq:ritter_simple}
\end{equation}

\noindent Here, $\rho_{L1}$ is the density at the $L_1$ point, $c_s$ is the sonic velocity at the photosphere, and $Q$ is the effective cross-section of the flow. The sonic velocity in the original Ritter formalism uses the ideal gas in a classic limit, with the assumption that the stream  evolves isothermally. 

In Figure~\ref{fig:mt_at}, we show the comparison of the original Ritter formalism {\bf OR} with the {\bf II,AI, AA} and {\bf EGA} models.
As expected, the difference between the $II$ and $IA$ cases is minimal, and confirms the expected ratio (see S~\ref{sec:atmooverflow}).

One would have expected that  {\bf OR} and {\bf II} would produce the same MT rate, but that is not what we find.
A significant part of the difference, a factor of few,  comes from the definition of the isothermal sonic velocity in our method, which is different from the one adopted in {\bf OR} ($\sqrt{P/\rho}$ in our case and $\sqrt{{\cal{R}}T/\mu}$ in {\bf OR} case, as they have specifically considered cold donors).
The MT rates in {\bf OR} are proportional to the sonic velocity to the power of 3, so our rates are always higher for hot donors, where the different values of isothermal sonic velocity can cause a difference in MT rates by a factor of 3-4. Only a linear proportionality can be easily seen in  Equation\ref{eq:ritter_simple}. The other two powers are hidden in the quantity $Q$. This quantity was originally proposed in \cite{1983A&A...121...29M}. It synthesizes several approximations, where the two most important ones are the Roche lobe approximation and the assumption on the density drop off with the distance from $L_1$ in units of pressure scale height. In the latter assumption, the sonic velocity implicitly enters, to the power of two. The remainder of the difference comes from the simplification of the Roche lobe geometry and the value of the pressure scale for the isothermal atmosphere; it is hard to disentangle the detailed role of each of the remaining approximation terms for comparing the used assumptions with the direct integrations.

Accounting for realistic atmospheric profiles for radiative donors results in higher mass loss rates in all atmospheric models as compared to {\bf OR} when the star is close to RLOF, with the {\bf EGA} and {\bf AA} models having lower MT rates when the donor is still far from the RLOF. In convective donors, the difference is smaller than in radiative donors for most atmospheric models. Usually, {\bf EGA} mass loss rates are slightly less than those predicted by {\bf OR} for the same fraction underflow.

\subsection{Unified formalism  in unperturbed donors}

\begin{figure*}
    \centering
    \includegraphics[width=0.5\textwidth]{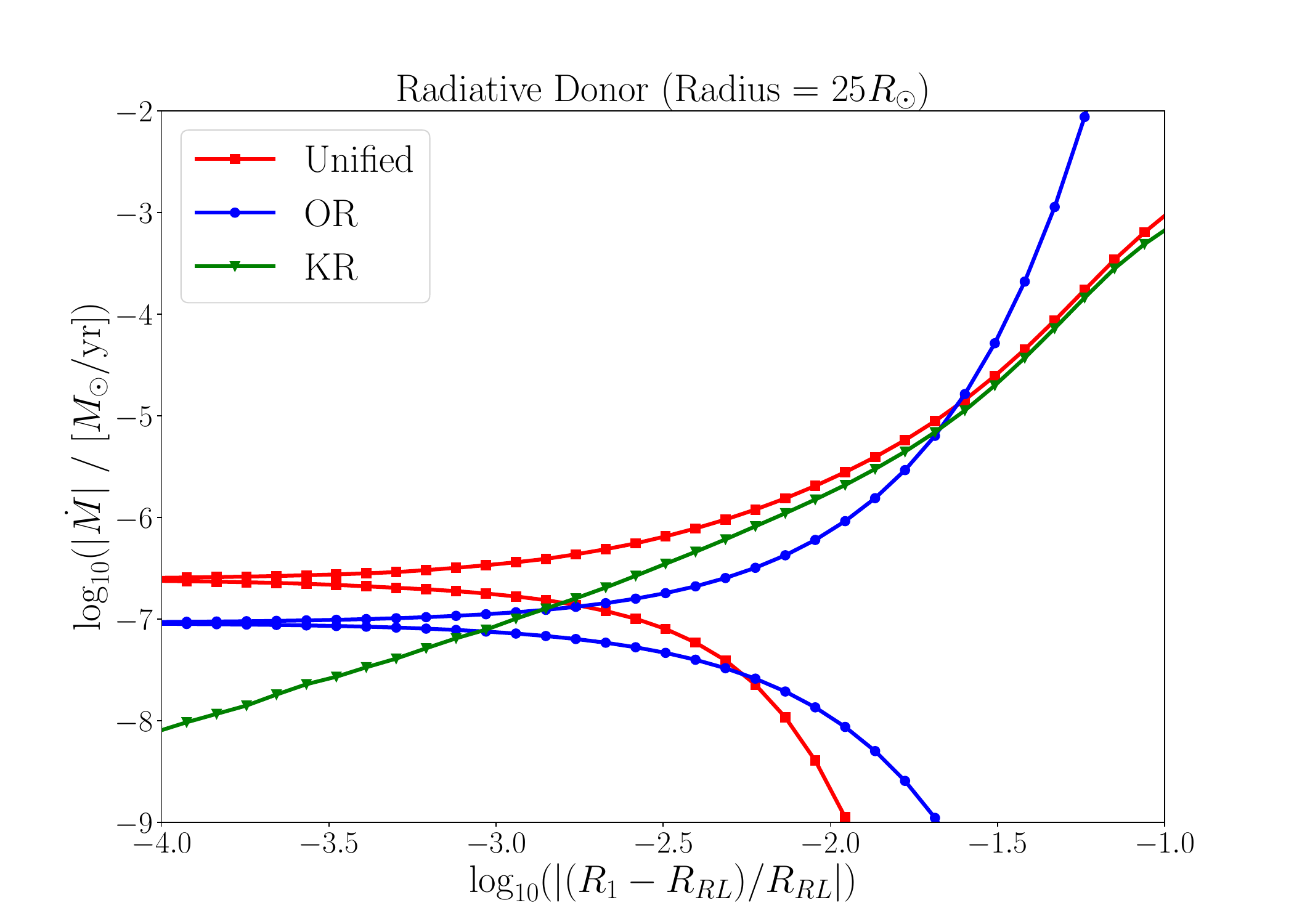}\includegraphics[width=0.5\textwidth]{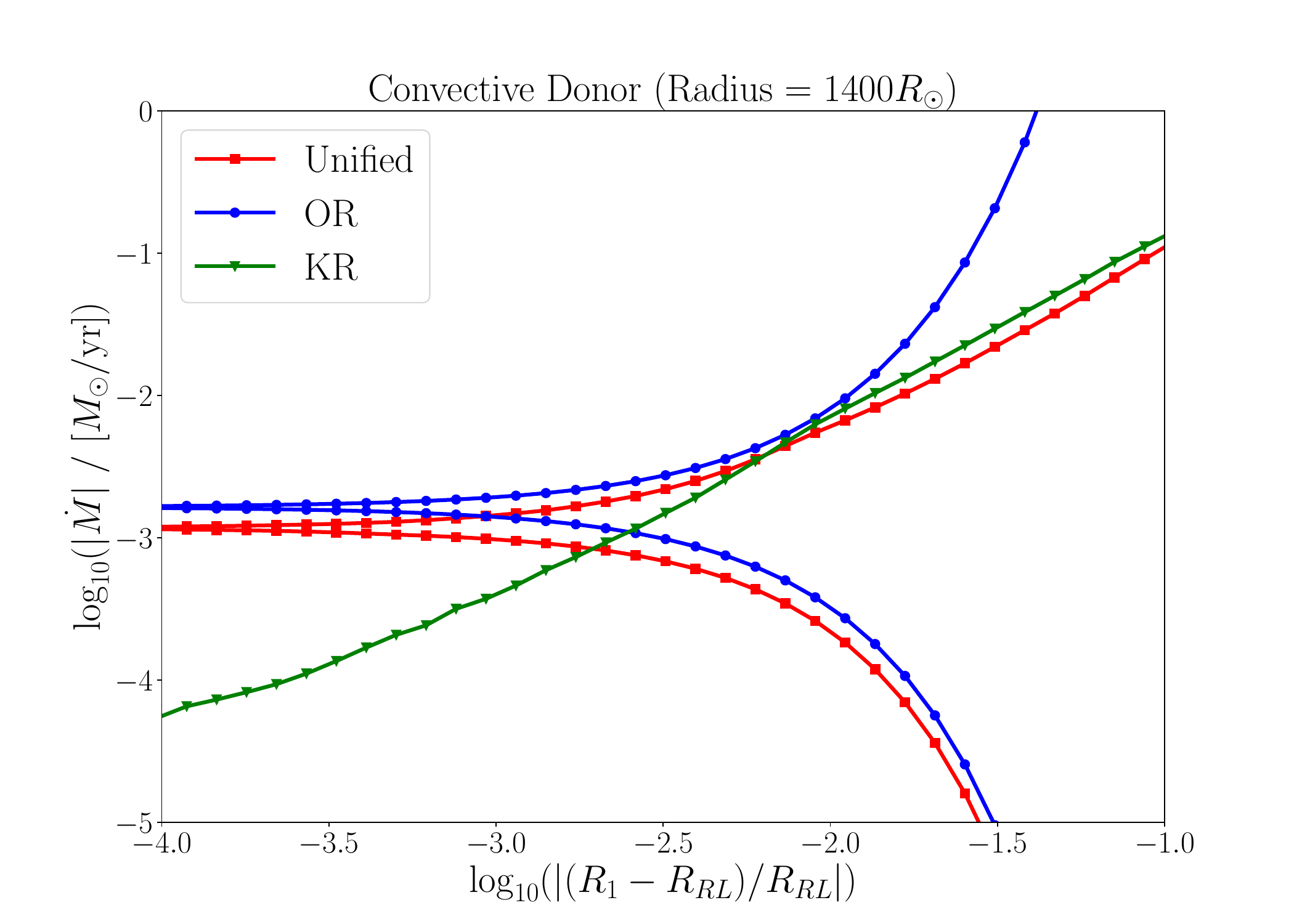}
    \caption{Mass loss rates as a function of distance to the Roche lobe. A radiative donor is shown in the left panel, and a convective donor is shown in the right panel. The models are described in \S\ref{sec:comp_whole}, specifically: ''Unified'' means our unified scheme without taking into account the effective binary potential, ``OR'' stands for {\bf OR} formalism, and ``KR'' stands for {\bf KR} formalism. {\bf KR} is shown only for the case of Roche overflow, while ''Unified'' and {\bf OR} are shown both for Roche underflow (atmospheric mass loss rate) and Roche overflow cases. The bottom branch in each case is for Roche underflow, and the upper branch is for Roche overflow.
        The initial mass of the donor is $30$~$M_\odot$, and the mass ratio $q=6$.}
    \label{fig:mt_nog}
\end{figure*}

\label{sec:comp_whole}

For this comparison, we select to use the {\bf EGA} atmospheric model for our unified model, although for very fast mass loss, the {\bf AA} atmospheric model is also acceptable.
In Figure~\ref{fig:mt_nog}, we show the comparison of our mass loss rate with the original Ritter formalism applied for both atmospheric overflow and for the donor's overflow, and optically thick mass loss rate using Kolb \& Ritter formalism \citep{1990A&A...236..385K}.
For this Figure, we found the MT rate obtained assuming that the same unperturbed donor produces mass loss as a function of the assumed underflow or overflow. No effective binary potential was used to modify the donor. For the comparison purposes of the prescriptions only during the binary evolution, we use default {\tt MESA} options except allowing for L-S coupling. If this option is permitted, the binaries of the same initial orbital separation may start the MT at different orbital separations. No direct comparison of the MT rate prescriptions  would then be possible.

As before, we find that for the considered radiative donor, the MT rates are lower in our formalism than in  {\bf OR} formalism when the donor is further away from the Roche lobe but becomes larger when the distance to the Roche lobe decreases. For convective donors, the difference is minimal. For overflow MT rates with very small overflow, the unified MT rates are larger than predicted by either Ritter or Kolb. For a more significant overflow, the unified MT rate is slightly larger in the radiative donor than by the Kolb prescription and is slightly smaller in the considered convective donor.

\section{Impact on the binary evolution}

\subsection{Effect of modified gravity} 

\begin{figure}[t]
	\centering                      
\includegraphics[width=\columnwidth]{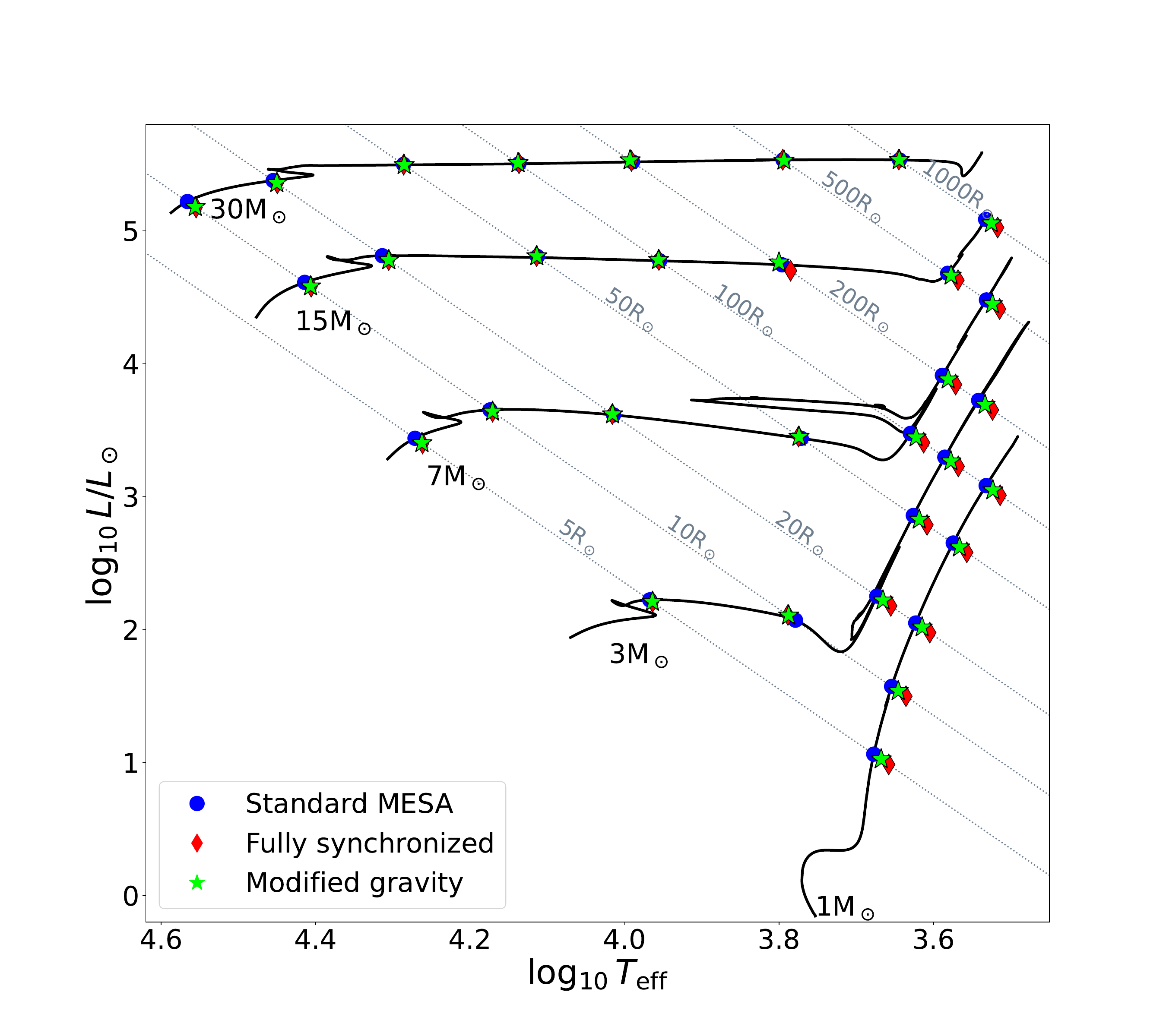}	
\caption{Comparison of $L$ and $T_{\rm eff}$ positions for  donors that have the same radius at the start of RLOF, but evolved using modified gravity (green stars), full synchronization (red diamonds), or with the standard {\tt MESA} binary package only (blue circles). These are shown for  binaries with a mass ratio of 10. Solid lines show the evolution of single stars of the given masses. Grey dotted lines correspond to the constant radii.}
	\label{fig:hrgrid}
\end{figure}

First, we considered the difference in the evolution before RLOF, with no atmospheric mass loss allowed, to capture only the effect of modified gravity. In addition to the already described standard binary evolution and our modified gravity evolution, we use {\tt MESA}'s standard binary package, including tidal evolution and orbital synchronization.
We consider five initial masses for the donor star: $ M_{\rm d}=1,3,7, 15$ and 30 $M_\odot$. For each mass, we pick up to 7 stellar models with the star's radius from the set of $5,10,20,50,100,200, 500$, and $1000 R_\odot$ (stars with some initial masses  never have either some small or some large radius values from this set). The calculations were done for several mass ratios between 1 and 10, but for the mass ratio $q=10$, the deviation between the evolutionary paths is most noticeable on the HR diagram, see Figure~\ref{fig:hrgrid}. 

We consider the cases of standard binary evolution and fully synchronized cases to be the two limiting cases, representing no rotation and the maximum allowed rotation. Our modified gravity case is expected to be close to the fully synchronized case (however, see Paper I, where we showed that the effective acceleration in a binary could differ from the solid rotation case by up to 20 percent). Overall, modified gravity by itself does not seem to affect the evolution of the donor significantly, and the change it provides is within the expected range. In rare cases, the modified gravity evolution path deviates slightly from being located precisely between the standard evolution and fully synchronized. It usually occurs when the donor transitions from having a radiative envelope to a convective envelope, as modified gravity causes a slightly different evolutionary path and envelope depths.

\begin{figure*}
    \centering
    \includegraphics[width=0.5\textwidth]{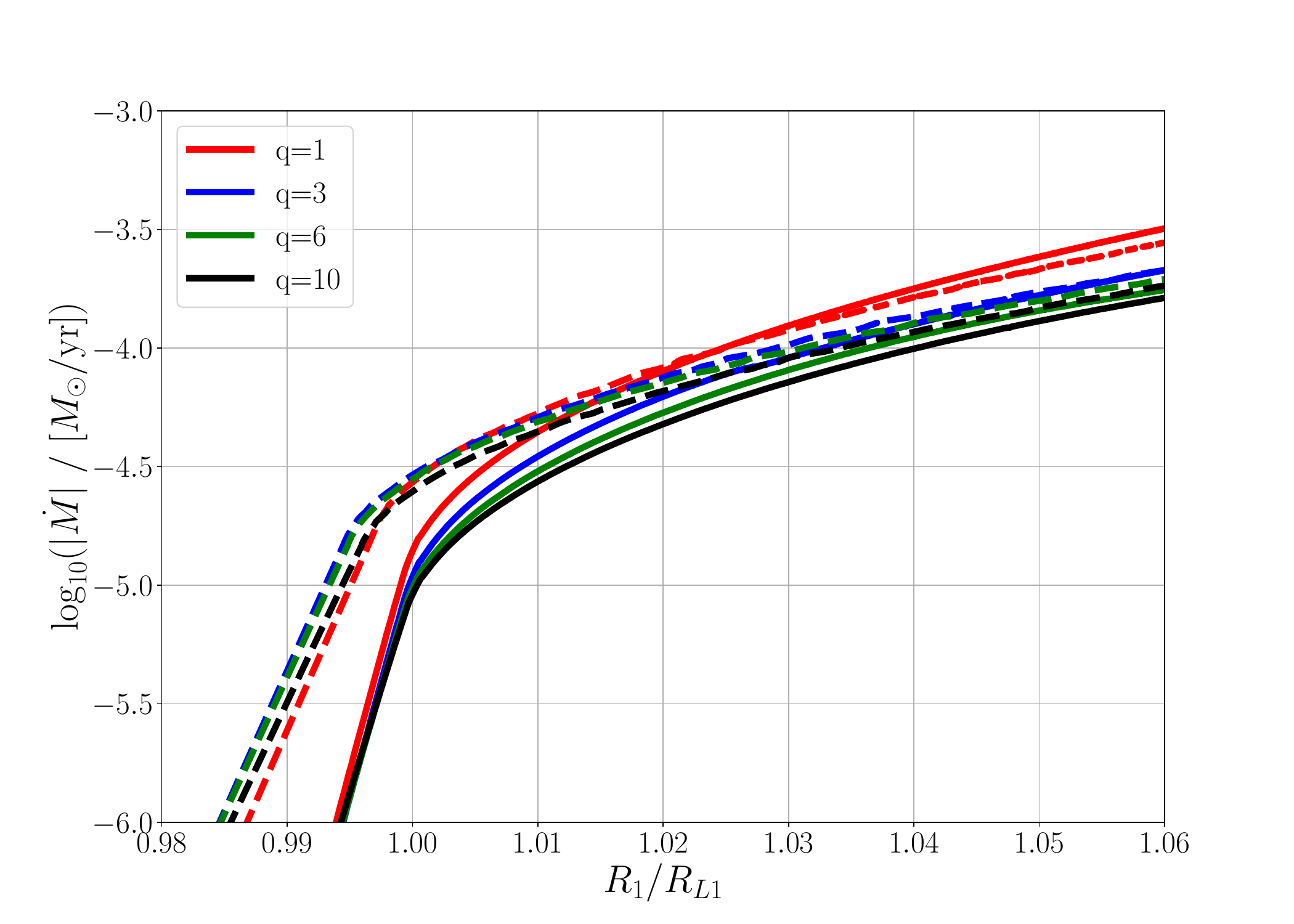}\includegraphics[width=0.5\textwidth]{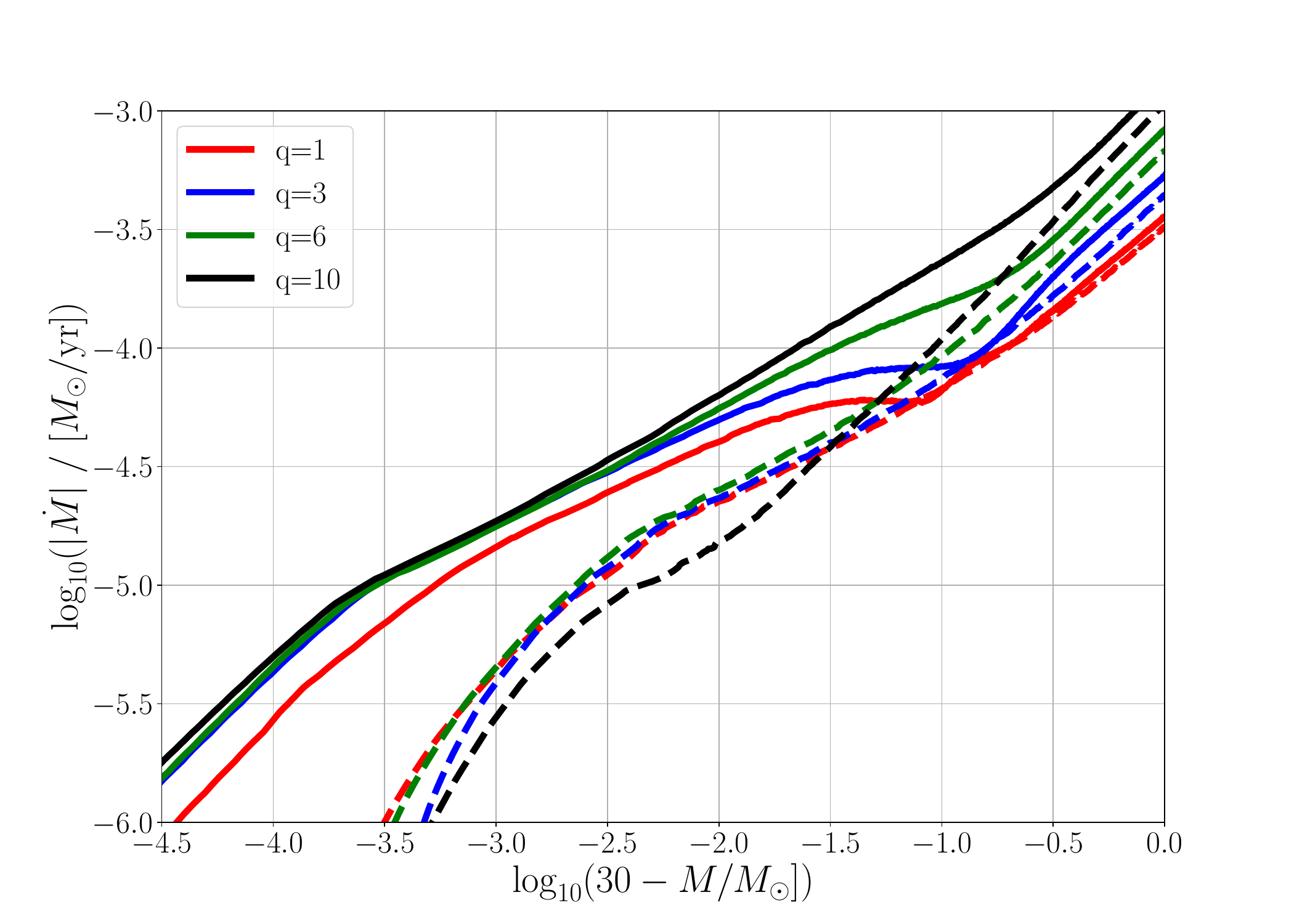}
    \caption{Mass loss rates as a function of the  Roche lobe overflow (the left panel) and the lost mass (the right panel) for a 30~$M_\odot$ donor in a binary system with  mass ratios $q=1,3,6$ and 10, and with such initial orbital separations that the radius of the donor is $R_1=$350~$R_\odot$ at RLOF. Solid lines show our method and dashed lines show the {\bf KR} prescription. In the left panel, we show behavior when the donor is close to RLOF. In the right panel,  we show only the evolution of MT rates during the time when the donor loses $1~M_\odot$.}
    \label{fig:mt_evol_qs}
\end{figure*}

\subsection{Unified mass loss formalism} 

We considered the set of donors with masses of $1,3,7,15$ and 30~$M_\odot$, the set of orbital separations from $a=10,20,40, 80, 160, 320, 640, 1280, 1800, 2400$~$R_\odot$, and the set of mass ratios $6$ and $10$ (and several cases with $q=1$ and $3$ and additional donor's radii at RLOF). We used only those orbital separations for each donor where the donor in the considered binary system starts RLOF. The comparison was made between standard {\tt MESA} binary setup with {\bf KR} mass loss and the same {\tt MESA} binary setup using our formalism while evolving the donor using  binary effective acceleration. 

We find that there are two kinds of MT endings. The first is when both formalisms ({\bf KR} and unified) lead to the overflow of the outer Lagrangian equipotential, and no sonic mass loss occurs. This situation is more typical for extended stars, although it also occurs in 1~$M_\odot$ donors with all considered separations. The comparison between the formalisms can be done either through the evolution of the MT episode, or for the amount of mass removed before $R> R_{\rm lout}$.

For example, in Figure~\ref{fig:mt_evol_qs}, we show the comparison of our revised rates and the {\bf KR} rates through the MT episode as a function of the RLOF and the lost mass.  The comparison is made for the donor of the same mass and radius at RLOF,  placed in binaries with different mass ratios.    We find that here atmospheric mass loss rates are always lower with our prescription for the same overflow, but they are higher for the same lost mass. The difference is especially noticeable, up ten times difference, for $q>1$. In the RLOF regime, our rates also consistently differ from those of {\bf KR}. In the shown mass transfer tracks, sonic mass loss $\dot M_{\rm sonic}$ was never achieved, and in all cases the donor has lost more than $1 M_\odot$. We note that the two ways to show tracks are not explicitly related, as the time spent at a specific overflow with a corresponding mass loss rate is an independent quantity, not shown.

\begin{figure*}
    \centering
    \includegraphics[width=0.49\textwidth]{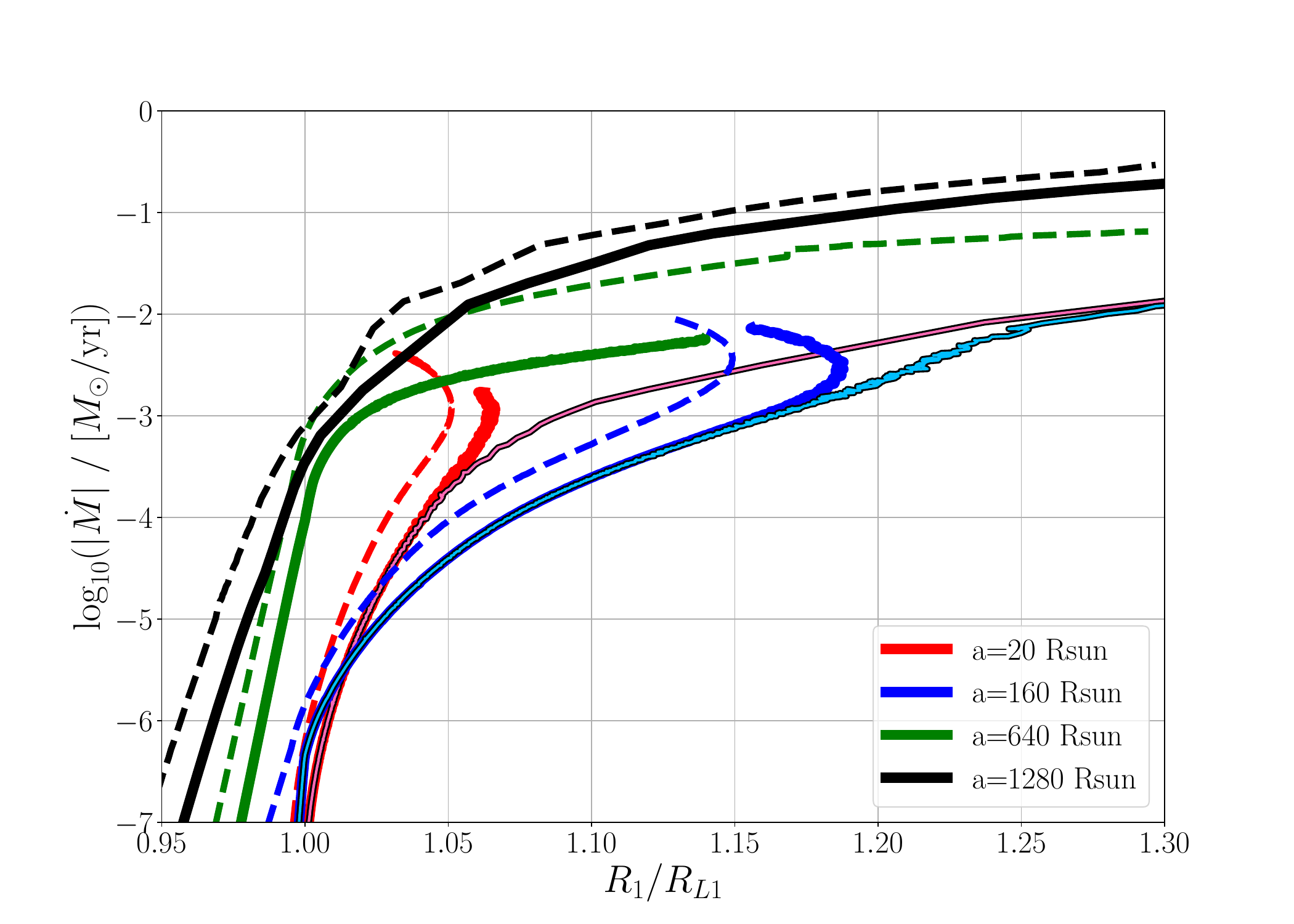}
    \includegraphics[width=0.49\textwidth]{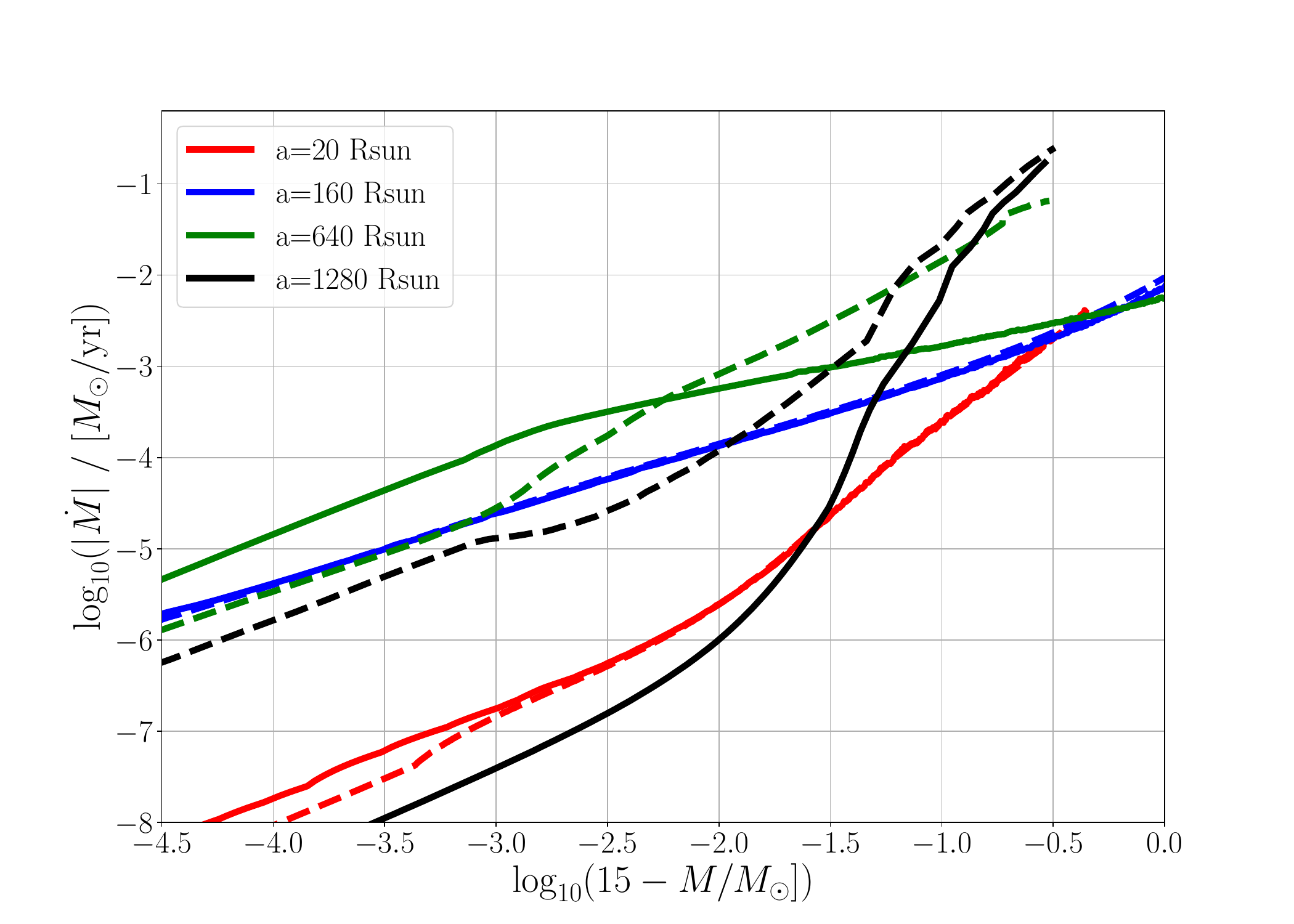}
    \caption{Mass loss rates as a function of the  Roche lobe overflow (left panel) and of the lost mass (right panel) for a $15 M_\odot$ donor in a binary system with  mass ratio $q=6$. The color indicates the initial orbital separation. Solid lines show our prescription, dashed lines show the {\bf KR} prescription. In both panels, we show only the evolution of MT rates while the donor loses the outermost $1~M_\odot$, where some binaries end their MT early. Black-bordered light red and light blue lines show the unified formalism with mass shell removal for $a=20~R_\odot$ and $a=160~R_\odot$.}
    \label{fig:mt_evol}
\end{figure*}

The second ending is when both formalisms lead (numerically) to MT exceeding the donor's sonic mass loss, sometimes to as high a MT as a factor of 100 times $\dot M_{\rm sonic}$.  The sonic situation occurs for stars close to their main sequence, and the reason for the code to stop is usually an inability to converge the donor's stellar model. In this case we can not compare the final (still converged) models, but can compare the MT rates before the sonic situation occurs.

For binaries where the MT rate exceeds the surface sonic mass loss, we do not find much difference in {$\dot M_{\rm tot}-$remaining mass tracks} during most of the RLOF mass transfer, although the responses to overflow ($\dot M_{\rm tot}-$overflow tracks) are different between the two formalisms. In Figure~\ref{fig:mt_evol}, we show the comparison of our revised MT rates and the {\bf KR} rate through the MT episode as a function of the lost mass, in the case of $M_{\rm d}=15 M_\odot$ and $q=6$, for several initial orbital separations. The sonic mass loss rates are achieved during the shown tracks for  binaries with the initial $a=20$~$R_\odot$, at $\log_{10} (\dot M_{\rm tot}/M_{\odot}$ yr$^{-1})=-3.8$, and $a=160$~$R_\odot$, at $\log_{10} (\dot M_{\rm tot}/M_{\odot}$ yr$^{-1})=-3.2$. It can be seen that when the outer layers are removed at a ''supersonic rate'', the donor starts quickly shrinking instead of continuing to expand; this effect is purely a numerical artifact. 

In verifying runs, where the predictive mass removal is switched on when $\dot M_{\rm tot} \ge 0.01 \dot M_{\rm sonic}$, the MT rates show the same response for most of the RLOF episode but provide a different $\dot M_{\rm tot}$ for the same remaining donor mass and RLOF when the $\dot M > 10 \dot M_{\rm sonic}$, see Figure~\ref{fig:mt_evol}. The predictive mass loss allows us to evolve to $R_1 = R_{\rm Lout}$, where the total mass lost by the end can be either smaller or larger than in the case of no predictive mass loss.  The $\dot M_{\rm tot}$-overflow tracks appear very similar to ''subsonic'' tracks in more expanded donors.

In the more extended donors, where the sonic mass loss rate is not achieved, the difference between the MT rates found using our formalism and the {\bf KR} formalism is similar to the cases shown in Figure~\ref{fig:mt_evol_qs}, and can be seen for $a=640$~$R_\odot$ and $a=1280$~$R_\odot$ in Figure~\ref{fig:mt_evol}.
Usually, the mass loss rates are higher for the same removed mass and lower for the same overflow.
The exception is the case of $a=$1280~$R_\odot$, where the combined effect of the evolution of the donor in a binary gravity field and the use of the {\rm EG$\Phi\kappa$} 
atmosphere, instead of isothermal, led to a lower rate of atmospheric outflow for a long period of time, resulting in a larger mass removed at a much lower mass loss rate. The RLOF for this binary starts when the mass loss rate is 
$\log_{10} (\dot M_{\rm tot}/M_{\odot}$ yr$^{-1}) \approx -3.2$ for our method  and $\log_{10} (\dot M_{\rm tot}/M_{\odot}$ yr$^{-1}) \approx -3$ for the {\bf KR} formalism. Switching the formalism leads to a difference of about 100 times in both the rate of atmospheric outflow and the time spent in outflow, for example with $\dot M > 10^{-8} M_{\odot}$ yr$^{-1}$, in the case of  unified mass loss, vs. the case of {\bf KR} formalism.

For the case with $a=640~R_\odot$, the RLOF starts at 
$\log_{10} (\dot M_{\rm tot}/M_{\odot}$ yr$^{-1})\approx -4.0$ for our formalism
and $\log_{10} (\dot M_{\rm tot}/M_{\odot}$ yr$^{-1})\approx -3.1$ for the {\bf KR} formalism. The main difference between the two formalisms in the stream mass loss rates (as a function of the overflow) and the outcomes for this donor is caused not by the MT prescription, but by evolving the donor using the effective binary potential. While only the evolution while losing the outermost $1 M_\odot$ is shown, the endings of the binaries are very different: the {\bf KR} prescription leads to  overflowing $R_{\rm lout}$ after losing $0.3 M_\odot$. In comparison, in the case of our unified prescription, almost the entire envelope is transferred ($9.1 M_\odot$) before the donor reaches $R_{\rm lout}$.
An opposite but similarly large difference is the case with the donor of $30 M_\odot$, $a=640 R_\odot$ and $q=10$: a donor evolved using our method can only transfer $\sim 1.1M_\odot$, but with the {\bf KR} method the donor lost $16.6 M_\odot$ before it reached $R_{\rm lout}$. 

Overall, we find that the new method always leads to the donor losing a different amount of mass before the onset of a CE event than if the same binary evolution model is used with the {\bf KR} method for MT. 
We note that the values of the mass lost in each case should be found by considering what the accretor is (an average star, a neutron star, or a black hole) and adopting the angular momentum loss mode appropriate to that specific binary. 

\section{Conclusion}

We present a revised method to obtain mass loss rates in 1D stellar codes. Our method aims to improve the treatment of mass loss at the onset of the MT episodes, especially in systems with a high mass ratio where MT evolves into an unstable regime. 

Our formalism uses the 3D average properties of the donor in a binary system for effective binary potentials and accelerations,
and can be applied in the range of mass ratios $10^{-2}\le q \le 10^2$.
The mass loss rates are found for stream MT and continuing atmospheric overflow simultaneously, integrating over the $L_1$ plane. For atmospheric overflow, we use atmospheric solutions obtained using opacities and gravitational acceleration as a function of the volume-equivalent radii during RLOF. To find the stream mass loss rate during RLOF, for the density and sonic velocity profiles of the stream at the $L_1$ plane, we use the donor's density and sonic velocity profiles from the same equipotentials, using both the donor interior and its atmosphere continuously. Our MT formalism is not restricted to cold donors. We also discussed how the mass has to be removed from the donor's stellar model in the case of a fast MT.

We find that our unified mass loss rates often differ by a factor of a few from those predicted by the \citet{Ritter1988} and \cite{1990A&A...236..385K} formalisms. Specifically, we find that our atmospheric mass loss rates, depending on the donor, can be larger by up to a factor of 10 or smaller by a factor of 100. This vastly different atmospheric mass loss may be important for binaries that evolve through long episodes of atmospheric overflow, like symbiotic binaries.
The difference between our method and the {\bf KR} method for overflow MT rates is smaller, and again, our rates can either exceed or be smaller than those predicted by the {\bf KR} formalism.
In some cases, the primary cause of the difference is the MT prescription. In other cases, it is the use of the binary effective potential. 
Overall, we find that our method removes a different mass before the moment the donor exceeds the outer Lagrangian equipotential.

The influence of the revised formalism on all existing mass-transferring systems (donor masses, mass ratios, orbital separations, donor and accretor types) is impossible to predict. The goal of this paper is to present the new formalism and test it against the {\bf OR}  and {\bf KR} formalisms using the same assumptions about binary evolution. Realistically, the adopted model of the angular momentum loss will play a significant role in the outcome. For example, mass losses from the accretor's outer Lagrangian point would speed up the system's shrinking, and the assumptions on how conservative the mass loss is also will change the outcomes. Simulations using appropriate models of angular momentum loss will be the subject of future studies of specific target groups of binaries. 

\begin{acknowledgments}
N.I. acknowledges funding from NSERC under Discovery Grant No. RGPIN-2019-04277. 
S.K. acknowledges funding from a MITACS Globalink Research Internship.
This research was enabled in part by support provided by Prairies DRI and the Digital Research Alliance of Canada (\url{alliancecan.ca}).
The authors thank an anonymous referee for their helpful comments that improved the quality of the manuscript.

\end{acknowledgments}

\vspace{5mm}
\facilities{Digital Research Alliance of Canada}

\bibliography{ref_rl}
\bibliographystyle{aasjournal}

\end{document}